\documentclass[doublespacing]{elsarticle}
\usepackage{graphicx}
\usepackage{amssymb,amsmath}

\usepackage{amssymb}
\usepackage[usenames,dvipsnames]{pstricks}
\usepackage{epsfig}

\begin{document}

\begin{frontmatter}

\title{Complex resonance frequencies of a finite, circular radiating duct with an infinite flange}
\author[LMA]{B. Mallaroni\corref{cor1}}
\cortext[cor1]{Corresponding author}
\ead{mallaroni@lma.cnrs-mrs.fr}
\author[LMA]{P.-O. Mattei}
\ead{mattei@lma.cnrs-mrs.fr}
\author[LMA]{J. Kergomard}
\ead{kergomard@lma.cnrs-mrs.fr}

\address[LMA]{Laboratoire de M\'ecanique et d'Acoustique, UPR CNRS 7051,
    31 chemin Joseph Aiguier, 13402 Marseille cedex 20, France}

\begin{abstract}
Radiation by solid or fluid bodies can be characterized by resonance
modes. They are complex, as well as resonance frequencies, because
of the energy loss due to radiation. For ducts, they can be computed
from the knowledge of the radiation impedance matrix. For the case
of a flanged duct of finite length radiating on one side in an
infinite medium, the expression of this matrix was given by
Zorumski, using a decomposition in duct modes. In order to calculate
the resonance frequencies, the formulation used in Zorumski's theory
must be modified as it is not valid for complex frequencies. The
analytical development of the Green's function in free space used by
Zorumski depends on the integrals of Bessel functions which become
divergent for complex frequencies. This paper proposes first a
development of the Green's function which is valid for all
frequencies. Results are applied to the calculation of the complex
resonance frequencies of a flanged duct, by using a formulation of
the internal pressure based upon cascade impedance matrices. Several
series of resonance modes are found, each series being shown to be
related to a dominant duct mode. Influence of higher order duct
modes and the results for several fluid densities is presented and
discussed.
\end{abstract}

\begin{keyword}
Acoustics \sep radiation impedance \sep cylindrical pipe \sep
resonance frequencies

\PACS 43.20.Rz \sep 43.20.Mv
\end{keyword}
\end{frontmatter}

\section{\label{sec:introduction}Introduction}

For the problem of duct radiation, many works have been done
concerning the calculation of radiation for a given duct mode, but
to the author's knowkedge no study has been done concerning the
resonance modes. We will treat the problem, in order to get a better
insight of the coupling of the duct and the surrounding space by
radiation. We choose the case of a duct with infinite flange,
because of its relative simplicity. One difficulty is due to the
fact that resonance frequencies (and modes) are complex, because
radiation is a form of dissipation. Notice that in the literature,
resonance modes are also called eigenmodes: \textit{they must be
distinguished from the duct modes} used in the present paper for the
purpose of the calculation.

 Green's functions are widely used in many
physical situations and notably in acoustics, for the calculation of
the pressure field radiated by physical sources (\emph{e.g.}
speakers, musical instruments, vibrating structures,...). For
instance, the solution given by Rayleigh \cite{rayleigh} to the
classical problem of a plane piston radiating into an infinite
flange involves the Green's function in free space. Zorumski \cite
{zorumski} extended this result to know the radiation of a
semi-infinite flanged duct in the form of matrix impedance, giving
the coupling between duct modes and used the Sonine's infinite
integral ( Ref. \cite{watson}, p.416, Eq.4) to develop the Green's
function in free space. In Refs \cite{shao1,kuijpers}, formulations
based on the Zorumski's method of this radiation impedance are
obtained for a larger class of problems. However, the development
fails when the frequency becomes complex: the corresponding infinite
integral, involving a Bessel function whose argument is a product of
the frequency and the dummy argument, becomes divergent for complex
frequencies. In many studies, the Zorumski's radiation matrix is
used as a boundary condition at the end of the duct in order to
calculate input impedances, length corrections or reflection
coefficients (see e.g. Refs. \cite{kergoamir2} or \cite{amir}).
 It is worth noting that complex resonance frequencies can occur in
various situations (\emph{e.g.} dissipative fluid, radiation,
complex impedance wall boundary conditions such as in Refs.
\cite{campos} or \cite{zorumski2}). In section \ref{sec:Green's
function} of this paper, we present a new expression for the Green's
function in free space for complex frequencies. In section
\ref{sec:section resonance frequencies}, an application of this
result is devoted to the determination of the complex resonance
frequencies of a cylindrical duct, closed at its input, considering
the influence of higher order duct modes. In the same secion, some
results are given and discussed. For this purpose, the internal
Green's function is previously calculated in section
\ref{sec:cascade impedance} with a method of cascade impedance.

\section{\label{sec:Green's function}Calculations of Green's function for the Helmholtz equation in free space}

Many studies on sound radiation by cylindrical ducts can be found in
the literature. For the case of an infinite flange, \ Norris and
Sheng \cite{norris} or Nomura \cite{namura} used a Green's function
integral to find an appropriate formulation for the external field.
We can also cite the classical work by Levine and Schwinger
\cite{levine} for the case of an unflanged pipe. Zorumski
\cite{zorumski}  extended the results for the planar mode to obtain
a multimodal radiation impedance which is a combination of the duct
modes present in the duct. In this section, these calculations are
briefly recalled, exhibiting the difficulty related to complex
frequencies. Thus, a new analytical formula for the Green's
function, valid for a dissipative problem, is presented.

\subsection{\label{sec:Zorumski's radiation impedance} Zorumski's radiation impedance}

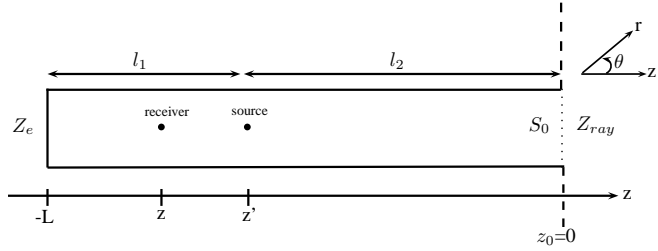
\begin{figure}[ht]
\begin{center}
\scalebox{0.75} 
{
\begin{pspicture}(0,-2.2117188)(11.690001,2.1917188)
\psline[linewidth=0.04cm](0.855625,0.6460937)(0.855625,-0.75390625)
\psline[linewidth=0.04cm](0.855625,-0.75390625)(10.015625,-0.73390627)
\psline[linewidth=0.04cm](0.835625,0.62609375)(9.955626,0.62609375)
\psline[linewidth=0.04cm](0.855625,-1.1339061)(0.855625,-1.4339062)
\psline[linewidth=0.04cm](2.8756251,-1.1339061)(2.8756251,-1.4339062)
\psline[linewidth=0.04cm](4.415625,-1.1339061)(4.415625,-1.4339062)
\psdots[dotsize=0.12](2.8756251,-0.03390625)
\psdots[dotsize=0.12](4.395625,-0.03390625)
\usefont{T1}{ptm}{m}{n}
\rput(0.800625,-1.6239064){-L}
\usefont{T1}{ptm}{m}{n}
\rput(2.8693752,-1.5839063){z}
\usefont{T1}{ptm}{m}{n}
\rput(4.4406247,-1.5839063){z'}
\usefont{T1}{ptm}{m}{n}
\rput(11.129375,-1.2239063){z}
\usefont{T1}{ptm}{m}{n}
\rput(0.42234376,-0.02390624){$Z_e$}
\usefont{T1}{ptm}{m}{n}
\rput(10.576407,-0.02390624){$Z_{ray}$}
\usefont{T1}{ptm}{m}{n}
\rput(2.971875,0.25109378){\scriptsize receiver}
\usefont{T1}{ptm}{m}{n}
\rput(4.440625,0.23109375){\scriptsize source}
\psline[linewidth=0.04cm,arrowsize=0.05291667cm
2.0,arrowlength=1.4,arrowinset=0.4]{<->}(0.855625,0.90609384)(4.275625,0.90609384)
\usefont{T1}{ptm}{m}{n}
\rput(2.5025,1.1360936){$l_1$}
\psline[linewidth=0.04cm,arrowsize=0.05291667cm
2.0,arrowlength=1.4,arrowinset=0.4]{<->}(4.335625,0.90609384)(9.955626,0.90609384)
\usefont{T1}{ptm}{m}{n}
\rput(7.0570316,1.1360936){$l_2$}
\psline[linewidth=0.04cm,arrowsize=0.05291667cm
2.0,arrowlength=1.4,arrowinset=0.4]{->}(0.15562499,-1.2539064)(10.975624,-1.2539064)
\psline[linewidth=0.03cm,linestyle=dotted,dotsep=0.10583334cm](9.975624,0.6460937)(9.975624,-0.75390625)
\usefont{T1}{ptm}{m}{n}
\rput(9.871718,-1.9839063){$z_0$=0}
\usefont{T1}{ptm}{m}{n}
\rput(9.573281,-0.00390625){$S_0$}
\psline[linewidth=0.03cm,arrowsize=0.05291667cm
2.0,arrowlength=1.4,arrowinset=0.4]{->}(10.317187,0.90488786)(11.477186,0.9057187)
\psline[linewidth=0.03cm,arrowsize=0.05291667cm
2.0,arrowlength=1.4,arrowinset=0.4]{->}(10.337188,0.9248879)(11.237187,1.6780833)
\usefont{T1}{ptm}{m}{n}
\rput(11.569375,0.93609375){z}
\usefont{T1}{ptm}{m}{n}
\rput(11.329062,1.7560937){r}
\rput{-67.61554}(5.6635094,10.545427){\psarc[linewidth=0.03,arrowsize=0.05291667cm
2.0,arrowlength=1.4,arrowinset=0.4]{->}(10.70573,1.0439299){0.13871127}{0.0}{165.59084}}
\usefont{T1}{ptm}{m}{n}
\rput(10.989063,1.1360937){$\theta$}
\psline[linewidth=0.04cm,linestyle=dashed,dash=0.16cm
0.16cm](9.96,0.63171875)(9.96,2.1717188)
\psline[linewidth=0.04cm,linestyle=dashed,dash=0.16cm
0.16cm](10.0,-0.74828124)(10.0,-1.8082813)
\end{pspicture}
}

\end{center}
\caption{\label{cylindre} Schema and coordinates of the duct.}
\end{figure}

We consider the radiation of sound into an infinite half space from
a circular duct (with radius $b$ and length $L$), with an infinite
flange at \textbf{$z_0 = 0$} (the index $0$ corresponds to the cross
section $S_0$ at the end of the duct) and we have chosen to work
with circular coordinates where the vector $\mathbf{r}$ is denoted
$(z,r,\theta)$ as shown by Fig. \ref{cylindre}.

The acoustic pressure in the infinite medium ($z\geq 0$) is given by
a Helmholtz integral (the time factor $\exp(-i\omega t)$ is omitted
throughout this paper):
\begin{equation}\label{pext}
p(\textbf{r})=-\frac{i\omega\rho}{2\pi}\int_0^{2\pi}\int_0^{b}r_0v(r_0,\theta_0)\frac{e^{ikh}}{h}dr_0d\theta_0,
\end{equation}
where $h=[r^2+r_0^2-2rr_0\cos(\theta-\theta_0)+z^2]^{\frac{1}{2}}$,
$\rho$ the ambient density and
 the wavenumber $k=\omega/c$ (with $\omega$ the circular frequency and $c$ the speed of sound).\\
The pressure $p$ and the velocity $v$ inside the duct ($z<0$) are
expressed as a series of duct eigen modes, so in $z=0$:
\begin{equation}\label{pzorum}
p(r,\theta,z=0)=\rho c^2\sum_m\sum_n\psi_{mn}(kr)e^{im\theta}P_{mn},
\end{equation}
\begin{equation}\label{vzorum}
v(r,\theta,z=0)=c\sum_m\sum_n\psi_{mn}(kr)e^{im\theta}V_{mn},
\end{equation}
where $\psi_{mn}(kr)e^{im\theta}$ is the transverse function for the
mode $mn$ with $\psi_{mn}(kr)=J_0(kr)/N_{mn}$. The $\lambda_{mn}$
are the eigenvalues, solutions of ${J}'(\lambda_{mn}kb)=0$, and the
norm $N_{mn}$
 is chosen similarly to that used by Zorumski \cite{zorumski}.\\
Substituting Eq. (\ref{vzorum}) into Eq. (\ref{pext}) gives the
pressure for $z\geq 0$ in terms of the modal velocity amplitudes
$V_{mn}$:
\begin{equation}\label{pextfinale}
p(r,\theta,z)=-\frac{i\omega\rho
c}{2\pi}\sum_m\sum_nV_{mn}\int_0^{2\pi}e^{im\theta_0}\int_0^{b}r_0\frac{e^{ikh}}{h}\psi_{mn}(kr_0)dr_0d\theta_0.
\end{equation}
Zorumski expressed the free space Green's function in equation
(\ref{pextfinale}) in terms of a Sonine's infinite integral
(\cite{watson}, p. 416, Eq. 4) and wrote for $z=0$ in the expression
of $h$:
\begin{equation}\label{sonine}
\frac{e^{ikh}}{h}=k\int_0^\infty\tau(\tau^2-1)^{-\frac{1}{2}}J_0(\tau
kh)d\tau.
\end{equation}
Next, he introduced a concept of "generalized radiation impedance
matrix $\mathbf{Z_{ray}}$" for a semi-infinite duct with an infinite
flange to describe the relation between the modal pressure and
velocity amplitudes:
\begin{equation}\label{Pmn}
P_{mn}=\sum_{l=1}^{\infty}Z_{mnl}V_{ml},
\end{equation}
where $m$, $l$ and $n$ are, respectively, the orders of
circumferential, radial incident and reflected modes. The element
$Z_{mnl}$ of the radiation impedance matrix gives the contribution
of the velocity mode $ml$ to the pressure mode $mn$ (for reasons of
symmetry, the coupling is possible only for a duct mode with the
same azimuthal dependance). The expression of the radiation
impedance is obtained as:
\begin{equation}\label{Zray_zorumski}
Z_{mnl}=-i\int_0^{\infty}\tau(\tau^2-1)^{-\frac{1}{2}}D_{mn}(\tau
,k)D_{ml}(\tau,k)d\tau,
\end{equation}
with, for a hard wall condition:
\begin{equation}\label{Dmn_zorumski}
D_{mn}(\tau,k)=kb\frac{\tau\psi_{mn}(kb)J_m'(\tau
kb)}{\lambda_{mn}^2-\tau ^2}.
\end{equation}

\subsection{\label{sec:Green's function in free space}Green's function for the Helmholtz equation in free space for complex frequencies}

The following asymptotic form (see Ref. \cite{abramowitz}, Eq.
9.2.1, p. 364) occurs when $\nu$ is fixed and $|\kappa|\rightarrow
\infty$:
\begin{equation}\label{abra}
J_{\nu}(\kappa)=\sqrt{\frac{2}{\pi
\kappa}}[cos(\kappa-\frac{1}{2}\nu \pi
-\frac{1}{4}\pi)+e^{|\Im(\kappa)|}O(|\kappa|^{-1})],
\end{equation}
with $|\arg \kappa|<\pi$ (in this paper, the real part and imaginary
part are represented, respectively, by the symbols $\Re$ and $\Im$).
As a consequence, for $\tau \rightarrow \infty$ with $k\in
\mathbb{C}$ we have $J_0(\tau kh) \rightarrow \infty$ when
$\Im(k)\neq0$, thus relation (\ref{sonine}) and the radiation
impedance (\ref{Zray_zorumski}) given by Zorumski are divergent
integrals for all non real
frequencies.\\
In order to have a Green's function for the Helmholtz equation in
free space valid for complex frequencies, we use another form of the
Sonine's infinite integral to develop this Green's function,
expressed by Watson \cite{watson} (p. 416, Eq. 4):
\begin{equation}\label{watson}
\frac{e^{ikh}}{h}=\int_0^{\infty}\tau(\tau^2-k^2)^{-\frac{1}{2}}J_0(\tau
h)d\tau.
\end{equation}
This integral remains convergent even for $k$ complex. A difficulty
occurs since $k$ is a branch point of the square root. For a time
factor $\exp(-i\omega t)$, the integration path on the real axis
must remain below $k$. However the complex resonance frequencies
$\omega=ck$ have a negative imaginary part (see explanation in
section \ref{sec:section resonance frequencies}), so the previous
formula must be adapted because of the branch cut. The integration
path below $k$ is classically deformed, as shown in Fig.
\ref{integ}:

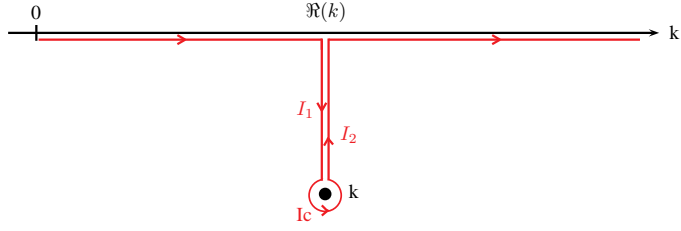
\begin{figure}[!ht]
\begin{center}
\scalebox{0.75} 
{
\begin{pspicture}(0,-2.2078125)(11.942187,1.9675)
\psdots[dotsize=0.6,linecolor=Red,fillstyle=solid,dotstyle=o](5.62,-1.4625)
\psline[linewidth=0.04cm,arrowsize=0.05291667cm
2.0,arrowlength=1.4,arrowinset=0.4]{->}(0.0,1.4175)(11.54,1.4175)
\usefont{T1}{ptm}{m}{n}
\rput(11.809531,1.4275){k}
\psline[linewidth=0.04cm](0.5,1.5575)(0.5,1.2775)
\usefont{T1}{ptm}{m}{n}
\rput(0.49703124,1.7875){0}
\psline[linewidth=0.04cm,linecolor=Red](0.54,1.2975)(5.56,1.2975)
\psline[linewidth=0.04cm,linecolor=Red](5.56,1.2775)(5.56,-1.2025)
\psline[linewidth=0.04cm,linecolor=Red](5.68,1.3175)(5.68,-1.1825)
\psline[linewidth=0.04cm,linecolor=Red](5.68,1.2975)(11.2,1.2975)
\usefont{T1}{ptm}{m}{n}
\rput(5.639375,1.7875){$\Re(k)$}
\psdots[dotsize=0.22399999](5.62,-1.4425)
\usefont{T1}{ptm}{m}{n}
\rput(6.1295314,-1.3925){k}
\psline[linewidth=0.04cm,linecolor=Red](3.14,1.2975)(3.0,1.3775)
\psline[linewidth=0.04cm,linecolor=Red](3.14,1.2975)(3.0,1.2175)
\psline[linewidth=0.04cm,linecolor=Red](5.56,0.0375)(5.48,0.1775)
\psline[linewidth=0.04cm,linecolor=Red](5.56,0.0375)(5.64,0.1775)
\psline[linewidth=0.04cm,linecolor=Red](5.68,-0.4425)(5.6,-0.5825)
\psline[linewidth=0.04cm,linecolor=Red](5.66,-0.4425)(5.76,-0.5625)
\psline[linewidth=0.04cm,linecolor=Red](5.66,-1.7425)(5.56,-1.6425)
\psline[linewidth=0.04cm,linecolor=Red](8.72,1.2975)(8.58,1.2175)
\psline[linewidth=0.04cm,linecolor=Red](8.72,1.2975)(8.58,1.3775)
\psdots[dotsize=0.08,linecolor=White](5.62,-1.1825)
\psline[linewidth=0.04cm,linecolor=White](4.34,-0.4225)(4.34,-0.5825)
\psline[linewidth=0.04cm,linecolor=White](3.72,-0.2425)(4.12,-0.3425)
\psline[linewidth=0.04cm,linecolor=White](4.12,-0.4425)(3.8,-0.7425)
\psline[linewidth=0.04cm,linecolor=Red](5.56,1.3175)(5.56,1.1175)
\usefont{T1}{ppl}{m}{n}
\rput(5.2703123,0.0475){\color{Red}$I_1$}
\usefont{T1}{ppl}{m}{n}
\rput(6.039375,-0.3925){\color{Red}$I_2$}
\usefont{T1}{ptm}{m}{n}
\rput(5.2495313,-1.8125){\color{Red}Ic}
\psline[linewidth=0.04cm,linecolor=Red](5.68,-1.7225)(5.56,-1.8225)
\end{pspicture}
} 
\end{center}
\caption{\label{integ} Deformation of the integration contour.}
\end{figure}

Now, the integral in Eq. (\ref{watson}) is written as:
\begin{equation}\label{watson_chemin}
\frac{e^{ikh}}{h}=\int_0^{|\Re(k)|}J_0(\tau
h)\tau(\tau^2-k^2)^{-\frac{1}{2}}d\tau+I_1(k,h)+I_{c}+I_2(k,h)+\int_{|\Re(k)|}^\infty
J_0(\tau h)\tau(\tau^2-k^2)^{-\frac{1}{2}}d\tau,
\end{equation}

with
$$I_1(k,h)=I_2(k,h)=-\int_{|\Re(k)|}^{k}{J}_0(\tau
h)\tau(\tau^2-k^2)^{-\frac{1}{2}}d\tau$$ and $I_c$ is zero
according to Jordan's lemma.\\
After calculations (similar to those developed by Morse and Feshbach
\cite{feshbach}, p.410), the following five cases can be
distinguished:
\\
i) $\Re(k)\leq0$ and $\Im(k)>0$
\begin{equation}\label{1}
\frac{e^{ikh}}{h}=\int_0^\infty J_0(\tau
h)\tau(\tau^2-k^2)^{-\frac{1}{2}}d\tau.
\end{equation}
\\
ii) $\Re(k)\leq0$ and $\Im(k)\leq0$
\begin{eqnarray}\label{2}
\frac{e^{ikh}}{h}&=&-i\int_0^{|\Re(k)|}J_0(\tau
h)\frac{\tau}{\sqrt{k^2-\tau^2}}d\tau+\int_{|\Re(k)|}^\infty
J_0(\tau
h)\frac{\tau}{\sqrt{\tau^2-k^2}}d\tau\nonumber\\
&&-2\int_{|\Re(k)|}^k J_0(\tau
h)\frac{\tau}{\sqrt{\tau^2-k^2}}d\tau.
\end{eqnarray}
\\
iii) $\Re(k)>0$ and $\Im(k)>0$
\begin{eqnarray}\label{3}
\frac{e^{ikh}}{h}=+i\int_0^{|\Re(k)|}J_0(\tau
h)\frac{\tau}{\sqrt{k^2-\tau^2}}d\tau +\int_{|\Re(k)|}^\infty
J_0(\tau h)\frac{\tau}{\sqrt{\tau^2-k^2}}d\tau.
\end{eqnarray}
\\
iv) $\Re(k)>0$ and $\Im(k)=0$
\begin{eqnarray}\label{4}
\frac{e^{ikh}}{h}&=&{J}_0(kh)k\sqrt{2\epsilon}(1-i)
+i\int_0^{|k|(1-\epsilon)}J_0(\tau
h)\frac{\tau}{\sqrt{k^2-\tau^2}}d\tau\nonumber\\
&&+\int_{|k|(1+\epsilon)}^\infty J_0(\tau
h)\frac{\tau}{\sqrt{\tau^2-k^2}}d\tau,
\end{eqnarray}
with $\epsilon << 1$.
\\
v) $\Re(k)>0$ and $\Im(k)<0$
\begin{eqnarray}\label{5}
\frac{e^{ikh}}{h}&=&+i\int_0^{|\Re(k)|}J_0(\tau
h)\frac{\tau}{\sqrt{k^2-\tau^2}}d\tau+\int_{|\Re(k)|}^\infty
J_0(\tau
h)\frac{\tau}{\sqrt{\tau^2-k^2}}d\tau\nonumber\\
&&-2\int_{|\Re(k)|}^k J_0(\tau
h)\frac{\tau}{\sqrt{\tau^2-k^2}}d\tau.
\end{eqnarray}

Contrary to the original Zorumski's formulation, these results
involve convergent integrals when the frequency is complex.

\subsection{\label{sec:New formulation of generalized impedance}New formulation of generalized impedance of a flanged
circular duct for real frequencies}

Initially, the previous result will be checked for the real case, in
order to compare the original Zorumski's result, noted in Eq.
(\ref{Zray_zorumski}), and that obtained using
   the expansion of $\exp(ikh)/h$ for the case (iv) where $\Im(k)=0$. Eq. (\ref{4}) leads to the following results:
\begin{eqnarray}\label{Zray_reelle}
Z_{mnl}&=&-ik[\tilde{D}_{mn}(k
)\tilde{D}_{ml}(k)k\sqrt{2\epsilon}(1-i)
+i\int_0^{|k|(1-\epsilon)}\frac{\tau}{\sqrt{k^2-\tau^2}}\tilde{D}_{mn}(\tau
)\tilde{D}_{ml}(\tau)d\tau\nonumber\\
&&+\int_{|k|(1+\epsilon)}^\infty\frac{\tau}{\sqrt{\tau^2-k^2}}\tilde{D}_{mn}(\tau
)\tilde{D}_{ml}(\tau)d\tau],
\end{eqnarray}
where
\begin{equation}\label{Dmn_moi}
\tilde{D}_{mn}(\tau)=b\frac{\tau\psi_{mn}(b)J_m'(\tau
b)}{\lambda_{mn}^2-\tau^2}.
\end{equation}

The radiation impedance for the planar mode ($m=n=0$ with $l=0$)
with Zorumski's formulation (\ref{Zray_zorumski}) and formulation
(\ref{Zray_reelle}) (with $\epsilon = 10^{-6}$) are very similarly
and thus, confirms the validity of
   formula (\ref{Zray_reelle}) with the identical computational cost.
 This formula is used in Ref. \cite{fab} to calculate an approximation of the reflection coefficient
  and of the length correction, taking into account the effect of the higher order duct modes below the first cut-off
  frequency.\\
The comparison between complex Zorumski's formulation for the planar
mode (m=n=0 and l=0) and Rayleigh's radiation impedance of a flanged
plane piston confirm the validity of $Z_{000}$ for all the
frequencies (see subsection \ref{sec:resonance frequencies without
HM}), because
 Rayleigh's radiation impedance is by definition the same quantity as $Z_{000}$.\\
In what follows, we show the interest of the radiation impedance
valid for complex frequencies when calculating the complex resonance
frequencies of a flanged finite length duct terminating in a
Zorumski's radiation condition. In a first instance, the internal
Green's function must be determined, with a method of cascade
impedances.

\section{\label{sec:cascade impedance}Calculation of the finite flanged duct internal Green's function with a method of cascade impedances}

\subsection{\label{sec:internal Green's function}Definition of the internal Green's function}

We search for the internal Green's function $G(M,M',\omega)$ at a
point $M(r,\theta,z)$ with a source in $M'(r',\theta',z')$,
satisfying:
\begin{equation}\label{Gverif}
(\Delta_M+k^2)G(M,M',\omega)=-\frac{1}{2\pi
r}\delta(r-r')\delta(\theta-\theta')\delta(z-z'),
\end{equation}
with $\partial_n G(M,M',\omega)=0$ on the walls. It is classically
(see e.g. Morse and Feshbach \cite{feshbach}) expanded in a series
of duct modes (the boundary conditions for the variable $z$ will be
given later on):
\begin{equation}\label{G}
G(M,M',\omega)=\sum^{\infty}_{m=0}\sum^{\infty}_{n=0}\phi_{mn}(r,\theta)\phi_{mn}(r',\theta
')g_{mn}(z,z',\omega),
\end{equation}
where $\phi_{mn}(r,\theta)=\psi_{mn}(r)e^{im\theta}$ is the
transverse function for the mode $mn$ and $g_{mn}(z,z',\omega)$ is
the longitudinal function for the mode $mn$. In what follows, the
dependance in $\omega$ of $g_{mn}(z,z',\omega)$ is omitted for
simplicity. It is worth noting that now we have the transverse
function with respect to $r$ and not $kr$, as per Zorumski
\cite{zorumski}. Therefore,
$\psi_{mn}(r)=\frac{J_m(\lambda_{mn}r)}{N_{mn}}$ where the
$\lambda_{mn}$ are solutions of ${J}'_m(\lambda_{mn} b)=0$, with
$\lambda_{mn}=\gamma_{mn}/b$. The $\gamma_{mn}$ are the $(n+1)^{th}$
zeros of the first derivative of the Bessel's function $J_m$. The
following norm $N_{mn}$ is chosen (Ref. \cite{zorumski} with respect
to $b$ and not $kb$):
\begin{equation}\label{norme_finale}
N_{mn}=b\sqrt{\pi}\sqrt{1-\frac{m^2}{\lambda^2_{mn}b^2}}J_m(\lambda_{mn}b).
\end{equation}

The transverse modes $\phi_{mn}(r,\theta)$ satisfy:
$$(\Delta_\bot+\lambda_{mn}^2)\phi_{mn}(r,\theta)=0,$$
with $\Delta_\bot=\frac{1}{r}\frac{\partial}{\partial
r}(r\frac{\partial}{\partial
r})+\frac{1}{r^2}\frac{\partial^2}{\partial \theta^2}$, thus:
$\Delta_\bot\phi_{mn}(r,\theta)=-\lambda_{mn}^2\phi_{mn}(r,\theta)$.
Studies such as Refs. \cite{kergoamir2} or \cite{amir} show that a
small finite number of terms is necessary for a numerical estimate
of the summation and it can be truncated: $M_m$ is defined as the
number of circumferential modes $m$ and $N_n$ as the number of
radial modes $n$. Introducing the previous expression (\ref{G}) in
(\ref{Gverif}) gives:
\begin{eqnarray}\label{G1}
\sum^{M_m}_{m=0}\sum^{N_n}_{n=0}(\frac{\partial^2}{\partial
z^2}+k^2_{mn})\phi_{mn}(r,\theta)\phi_{mn}(r',\theta ')g_{mn}(z,z')
=-\frac{1}{2\pi r}\delta(r-r')\delta(\theta-\theta')\delta(z-z'),
\end{eqnarray}
where $k^2_{mn}=k^2-\lambda^2_{mn}$. Then, multiplying the left and
right sides of (\ref{G1}) by $\phi_{m'l}^*(r,\theta)$ and
integrating the resulting equality on the surface S, the
orthogonality condition
($\int_S\phi_{mn}\phi_{m'l}^*dS=\delta_{mm'}\delta_{nl}$), leads to:
\begin{equation}\label{relation_P}
(\frac{\partial^2}{\partial
z^2}+k^2_{mn})g_{mn}(z,z')=-\delta(z-z').
\end{equation}

With a conservative Neumann or Dirichlet boundary condition at the
extremities of the duct, the resonance wavenumbers can be easily
computed. But with a dissipative boundary condition like those
occurring for the sound radiation of a flanged cylinder, there is no
simple analytical solution. Therefore, in the next section, the
elements $g_{mn}(z,z')$ are calculated with a method of cascade
impedances presented in Refs. \cite{kergoamir2},\cite{kergoamir1} or
\cite{kergo_transmi}.

\subsection{\label{sec:method of cascade impedances}Presentation of the method of cascade impedances}

The calculation of the duct impedance at an abscissa $z_1$ with
respect to another abscissa $z_2$ is based on the following transfer
matrix relationship (see e.g. Ref. \cite{kergo_transmi}):

\begin{equation}
\label{trans} \left(
\begin{array}{c}
P_{mn}(z_1) \\
V_{mn}(z_1)
\end{array}
\right) = \left( M_T \right) \left(
\begin{array}{c}
P_{mn}(z_2) \\
V_{mn}(z_2)
\end{array}
\right),
\end{equation}

$M_T=\left(
\begin{array}{cc}
\cosh(ik_{mn}(z_2-z_1)) & Z_{c,mn}\sinh(ik_{mn}(z_2-z_1)) \\
\displaystyle \frac{\sinh(ik_{mn}(z_2-z_1))}{Z_{c,mn}} &
\cosh(ik_{mn}(z_2-z_1))
\end{array}
\right)$, where $ Z_{c,mn} $ is an element of the diagonal matrix of
characteristic impedance $\mathbf{Z_c}$:
$$ Z_{c,mn} = \frac{k\rho c}{k_{mn}}.$$
Moreover, matrices formulation is chosen: $\mathbf{\Phi}(r,\theta)$
is a column vector constituted by the $M_mN_n$ elements $\phi_{mn}$,
verifying
$\int_S\mathbf{\Phi}(r,\theta)\mathbf{\Phi}^{T}(r,\theta)dS=\mathbf{I}$,
$\mathbf{P}(z)$ is a column vector constituted by the $M_mN_n$
elements $P_{mn}$ and $\mathbf{V}(z)$ is a column vector constituted
by the $M_mN_n$ elements $V_{mn}$. Thus, relation (\ref{trans}) is
now written:
\begin{equation}
\label{trans_matrice} \left(
\begin{array}{c}
\mathbf{P}(z_1) \\
\mathbf{V}(z_1)
\end{array}
\right) = \left(
\begin{array}{cc}
\mathbf{C} & \mathbf{Z_{c}S} \\
\displaystyle \mathbf{Z_{c}^{-1}S} & \mathbf{C}
\end{array}
\right) \left(
\begin{array}{c}
\mathbf{P}(z_2) \\
\mathbf{V}(z_2)
\end{array}
\right),
\end{equation}
where $\mathbf{C}$ and $\mathbf{S}$ are diagonal matrices
constituted by the elements $\cosh(ik_{mn}(z_2-z_1))$ and
$\sinh(ik_{mn}(z_2-z_1))$, respectively.
\\
The transfer matrix formulation (\ref{trans_matrice}) is now
transformed into an impedance matrix formulation. The calculation,
presented in Ref. \cite{kergo_transmi}, is recalled in Appendix A.
This gives the following matrix equation:
\begin{equation}
\label{multiporte} \left(
\begin{array}{c}
\mathbf{P}(z_1) \\
\mathbf{P}(z_2)
\end{array}
\right) = \left(
\begin{array}{cc}
\mathbf{Z_{11}} & -\mathbf{Z_{12}} \\
\displaystyle \mathbf{Z_{21}} & -\mathbf{Z_{22}}
\end{array}
\right) \left(
\begin{array}{c}
\mathbf{V}(z_1) \\
\mathbf{V}(z_2)
\end{array}
\right),
\end{equation}

where $\mathbf{Z_{11}}=\mathbf{Z_cS^{-1}C}$,
$\mathbf{Z_{12}}=\mathbf{Z_cS^{-1}}$,
$\mathbf{Z_{21}}=\mathbf{Z_cS^{-1}}$ and
$\mathbf{Z_{22}}=\mathbf{Z_cS^{-1}C}$.

\subsection{\label{sec:Pz'}Calculation of \textbf{P}(z')}

First, the pressure vector at the source position $z'$ is
calculated. For this purpose, we calculate a right-side matrix
impedance $\mathbf{Z}^{+}(z')$ at $z'$ with respect to
$\mathbf{Z_{ray}}$, a left-side matrix impedance
$\mathbf{Z}^{-}(z')$ at $z'$ with respect to $\mathbf{Z}_{e}$ and
finally the connection
 between these two matrices at $z'$, the abscissa of the source.\\

\textbf{Step 1: Right-side matrix impedance
$\mathbf{Z^{+}(z')}$}\\
Let us denote $\mathbf{C_s}$ and $\mathbf{S_s}$ the diagonal matrix
constituted by the elements $\cosh(ik_{mn}l_2)$
 and $\sinh(ik_{mn}l_2)$, respectively, where $l_2$ (see Fig. \ref{cylindre}) is the distance
 between the abscissa $z'$ and the extremity ($z=0$). The radiation impedance
 $\mathbf{Z_{ray}}$, constituted by the elements $Z_{mnl}$ calculated in the previous section, verifies
 $\mathbf{P}(z'+l_2)=\mathbf{Z_{ray}V}(z'+l_2)$. Thus Eq. (\ref{trans_matrice}) leads to:
\begin{equation}\label{B}
\mathbf{V}(z'+l_2)=(\mathbf{Z_{ray}+\mathbf{Z_{22}}})^{-1}\mathbf{Z_{21}V}(z'),
\end{equation}
and, with Eqs. (\ref{multiporte}) and (\ref{B}), to:
$$\mathbf{P}(z')=\mathbf{Z_{11}V}(z')-\mathbf{Z_{12}}(\mathbf{Z_{ray}}+\mathbf{Z_{22}})^{-1}\mathbf{Z_{21}V}(z').$$
With $\mathbf{Z^+}$ verifying
$\mathbf{P}(z')=\mathbf{Z^+}(z')\mathbf{V}(z')$, we have finally:
\begin{equation}\label{Zplus0}
\mathbf{Z^+}(z')
=\mathbf{Z_{11}}-\mathbf{Z_{12}}(\mathbf{Z_{ray}}+\mathbf{Z_{22}})^{-1}\mathbf{Z_{21}},
\end{equation}
or:
\begin{equation}\label{Zplus}
\mathbf{Z^+}(z')=\mathbf{Z_cS_s^{-1}C_s}-\mathbf{Z_cS_s^{-1}}[\mathbf{Z_c^{-1}Z_{ray}}+\mathbf{S_s^{-1}C_s}]^{-1}\mathbf{S_s^{-1}}.
\end{equation}
\\

\textbf{Step 2: Left-side matrix
impedance $\mathbf{Z^{-}(z')}$}\\
 We choose a Neumann condition for $z=-L$ (thus $\mathbf{V}(z'-l_1)=0$).
 Here, $\mathbf{C_e}$ and $\mathbf{S_e}$ are the diagonal matrices constituted by the
elements $\cosh(ik_{mn}l_1)$ and $\sinh(ik_{mn}l_1)$, respectively.
$l_1$ (see Fig. \ref{cylindre}) is the
 distance between the point $z=-L$ and the point $z'$. The impedance $\mathbf{Z_{e}}$ is calculated at $z=-L$.\\
Relation (\ref{multiporte}) is written for the present case as:
\begin{equation}\label{Pz0l2}
\mathbf{P}(z')=\mathbf{Z_cS_e^{-1}}\mathbf{V}(z'-l_1)-\mathbf{Z_cS_e^{-1}}\mathbf{C_e}\mathbf{V}(z').
\end{equation}
Thus, with the Neumann condition in $z=-L$ and with Eq.
(\ref{Pz0l2}), using
$\mathbf{P}(z')=\mathbf{Z^-}(z')\mathbf{V}(z')$, the following
result is obtained:
\begin{equation}\label{Zmoins}
\mathbf{Z^-}(z')=-\mathbf{Z_cS_e^{-1}C_e}.
\end{equation}

\textbf{Step 3: Connection between the impedance matrices $\mathbf{Z^+}$ and $\mathbf{Z^-}$ at $\mathbf{z'}$}\\
Let us denote $P^{\pm}_{mn}(z')=[g_{mn}(z,z')]_{z=z'\pm \epsilon}$.
Using the continuity of the Green's function at $z=z'$, leads to
when $\epsilon \rightarrow 0$:
\begin{equation}\label{continu}
P^+_{mn}(z')=P^-_{mn}(z')=P_{mn}(z'),
\end{equation}
Integrating relation (\ref{relation_P}) on an interval of width
$2\varepsilon$  between $z'+\varepsilon$ and $z'-\varepsilon$ gives:
\begin{equation}\label{eps2}
\int^{z'+\varepsilon}_{z'-\varepsilon}(\frac{\partial^2}{\partial_{z^2}}+k^2_{mn})g_{mn}(z,z')dz=-1
\end{equation}
and, with the pressure continuity, when $\varepsilon \rightarrow 0$:
\begin{equation}\label{sautderiv}
\partial_{z}P^+_{mn}(z')-\partial_{z}P^-_{mn}(z')=-1,
\end{equation}
with
$\partial_{z}P^{\pm}_{mn}(z')=[\partial_{z}g_{mn}(z,z')]_{z=z'\pm
\epsilon}$.\\
Euler's dimensionless equation $\frac{1}{\rho
c}V_{mn}(z')=-\frac{i}{\omega\rho}\partial_{z}P_{mn}(z')$ implies
$\partial_{z}P_{mn}(z')={ik}V_{mn}(z')$, thus, using Eq.
(\ref{sautderiv}):
\begin{equation}\label{sautU}
V^+_{mn}(z')-V^-_{mn}(z')=-\frac{1}{ik}.
\end{equation}
We introduce a column vector $\mathbf{W}$ of $M_mN_n$ lines whose
elements equal $1$. Equation (\ref{sautU}) may be expressed as:
$$(\mathbf{Z^+}(z'))^{-1}\mathbf{P^+}(z')-(\mathbf{Z^-}(z'))^{-1}\mathbf{P^-}(z')=-\frac{1}{ik}\mathbf{W},$$
and using Eq. (\ref{continu}), the pressure at the source is written
as follows:
\begin{equation}\label{Pz}
\mathbf{P}(z')=-\frac{1}{ik}\left[(\mathbf{Z^+}(z'))^{-1}-(\mathbf{Z^-}(z'))^{-1}\right]^{-1}\mathbf{W}.
\end{equation}

\subsection{\label{sec:gzz'}Expression of the function \textbf{g}(z,z')}
We introduce a column vector $\mathbf{g}(z,z')$ constituted by the
$M_mN_n$ elements $g_{mn}(z,z')$. They are two possible
configurations with respect to the relative
positions of receiver at $z$ and source at $z'$:\\

\textbf{First configuration: $\mathbf{z>z'}$}\\
Let $l_r=z-z'$ be the distance between the receiver and the source
(see Fig. \ref{cylindre}), $\mathbf{C_{l_r}}$ the diagonal matrix
constituted by the elements $\cosh(ik_{mn}l_r)$, and
 $\mathbf{S_{l_r}}$ the diagonal matrix constituted by the
elements $\sinh(ik_{mn}l_r)$. Relation (\ref{trans}) gives for
$z>z'$:
$$\mathbf{g}(z,z')=\mathbf{C_{l_r}P}(z')-\mathbf{Z_cS_{l_r}}\mathbf{V}(z'),$$
then, with $\mathbf{P}(z')=\mathbf{Z^+}(z')\mathbf{V}(z')$:
\begin{equation}\label{Pz}
\mathbf{g}(z,z')=\mathbf{S_{l_r}}\left[\mathbf{S_{l_r}^{-1}C_{l_r}}-\mathbf{Z_c}(\mathbf{Z^+}(z'))^{-1}\right]\mathbf{P}(z').
\end{equation}

\textbf{Second configuration: $\mathbf{z'>z}$}\\
Let $l_l=z'-z$ be the distance between the receiver and the
source(see Fig. \ref{cylindre}), $\mathbf{C_{l_l}}$ the diagonal
matrix constituted by the elements $\cosh(ik_{mn}l_l)$, and
 $\mathbf{S_{l_l}}$ the diagonal matrix constituted by the
elements $\sinh(ik_{mn}l_l)$. Relation (\ref{trans}) gives for
$z'>z$:
$$\mathbf{g}(z,z')=\mathbf{C_{l_l}}\mathbf{P}(z')+\mathbf{Z_c}\mathbf{S_{l_l}}\mathbf{V}(z'),$$
then, with $\mathbf{P}(z')=\mathbf{Z^-}(z')\mathbf{V}(z')$:
\begin{equation}\label{Pzg}
\mathbf{g}(z,z')=\mathbf{S_{l_l}}\left[\mathbf{S_{l_l}^{-1}C_{l_l}}+\mathbf{Z_c}(\mathbf{Z^-}(z'))^{-1}\right]\mathbf{P}(z').
\end{equation}
Finally, the Green's function of a finite duct with an infinite
flange is given as:
\begin{equation}\label{internal_green}
\mathbf{G}(M,M')=\mathbf{\Phi}(r,\theta)\mathbf{\Phi}(r',\theta')\mathbf{g}(z,z').
\end{equation}

\section{ \label{sec:section resonance frequencies} Application to complex resonance frequencies of a flanged,
finite length duct} Resonances of a flanged, finite length duct are
interesting as they contain important information about the coupling
between internal and external fluids.
    Their calculation is based on the fact that the internal pressure becomes infinite at each resonance.
    Newton's method is used to compute
    the zeros of the inverse of the pressure. Since the resonances of a dissipative problem are complex,
     a complex formulation of the impedance radiation is needed.
As a time dependence $\exp(-i\omega t)$ has been chosen, the
imaginary part needs to be negative for resonance frequencies in
order to ensure that the amplitude remains bounded for all times
$t$.
 Using the integrals (\ref{2}) and (\ref{5}), for $\Re(k)>0$ and $\Im(k)<0$, $Z_{mnl}$ becomes:
\begin{eqnarray}\label{Zray_complexe}
Z_{mnl}&=&-ik[i\int_0^{|\Re(k)|}\frac{\tau}{\sqrt{k^2-\tau^2}}\tilde{D}_{mn}(\tau
)\tilde{D}_{ml}(\tau)d\tau
+\int_{|\Re(k)|}^{+\infty}\frac{\tau}{\sqrt{\tau^2-k^2}}\tilde{D}_{mn}(\tau
)\tilde{D}_{ml}(\tau)d\tau\nonumber\\
&&-2\int_{|\Re(k)|}^{k}\frac{\tau}{\sqrt{\tau^2-k^2}}\tilde{D}_{mn}(\tau
)\tilde{D}_{ml}(\tau)d\tau].
\end{eqnarray}
This expression is used as a radiation condition $\mathbf{Z_{ray}}$
at the end of the finite length duct.\\

\subsection{\label{sec:resonance frequencies without HM}Resonance wavenumbers for the planar mode without influence of higher order duct modes}

In a first instance, we consider the resonance wavenumbers of the
planar duct mode (m=n=0) without the influence of higher order duct
modes (l=0). With radiation, the $j^{th}$ resonance wavenumbers of
duct mode $mn$ are denoted $k_{mn,r}^{j}$ and denoted
$k_{mn}^{j}$ without radiation.\\
In order to validate the complex formulation of the radiation
impedance, we compare the resonances obtained with radiation
impedance given by relation (\ref{Zray_complexe}) for $m=n=0$ and
$l=0$, with the resonances obtained with the radiation impedance of
a flanged plane piston given by Rayleigh's formulation as (see Ref.
\cite{feshbach} p.1458):
\begin{equation}\label{rayleigh}
Z_0\simeq\rho c[1-\frac{1}{kb}J_1(2kb)-\frac{i}{kb}S_1(2kb)]
\end{equation}

\begin{figure}[!ht]
\begin{center}
\includegraphics[height=.25\textheight,width=.6\linewidth]{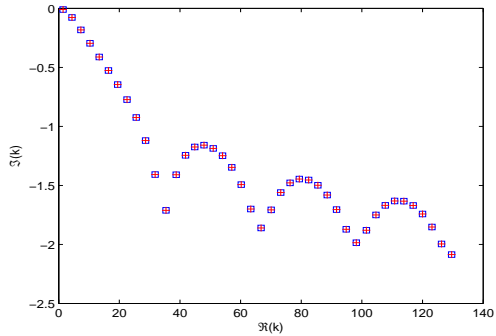}
\end{center}
\caption{\label{zero_pole_000} Evolution of resonance wavenumbers
$k_{00,r}^j$ depending on wavenumber $k$ (with $b=0.1m$ and $L=1m$)
with the radiation impedance defined by Eq. (\ref{Zray_complexe})
for $m=n=0$ and $l=0$ ($+$) and with the radiation impedance of a
flanged plane piston (Eq. \ref{rayleigh}) ($\Box$).}
\end{figure}

The radiation impedance for m=n=l=0 calculated with relation
(\ref{Zray_complexe}) and that of a flanged plane piston
(\ref{rayleigh}) are identical. Thus, resonance wavenumbers
calculated with these two formulations are so identical, as observed
in Fig. \ref{zero_pole_000}.

It is worth noting in Table \ref{table1} that without radiation, the
resonance frequencies are those obtained by the usual longitudinal
resonances of a cylindrical duct with one side "closed" and the
other side "open" (Neumann/Dirichlet problem) for:
$$k_{00}^j=\frac{(2j+1)\pi}{2L},j=0,1,2,...$$
We can observe in Table \ref{table1} of Appendix B that the real
part of resonance frequency decreases when radiation is taken into
account: this is normal behavior because the reactive effect of
radiation can roughly be described as an increase in the duct
length. The semi-infinite duct length correction is estimated by the
following formula (see Ref. \cite{norris} with the modification
given in private communication):
\begin{equation}\label{corec_long}
\frac{\triangle L}{b}
=0.82159\frac{1+\frac{kb}{1.2949}}{1+\frac{kb}{1.2949}+(\frac{kb}{1.2949})^2}.
\end{equation}
The difference between the real part of the $j$ first resonance
wavenumbers $k_{00,r}^{j}$ of a finite radiating cylindrical duct
(below the first cut off wavenumber $k_{cut01}=3.83/b$) and those
estimated using the length correction defined by Eq.
(\ref{corec_long}) ($k_{00,\triangle L}^{j}=(2j+1)\pi/[2(L+\triangle
L)]$),
  tends to zero when the ratio $L/b$ increases. The values agree well
  ($\leq 1\%$ of difference) for $L/b \geq 3$ with several values of $L$.

\subsection{Resonance wavenumbers with influence of higher order duct modes}
\begin{figure}[!ht]
\begin{center}
\includegraphics[height=.25\textheight,width=.6\linewidth]{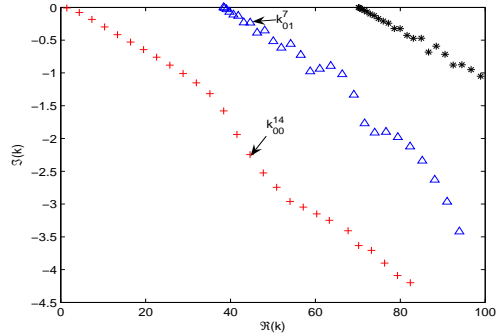}
\end{center}
\caption{\label{zero_3HM}Resonance wavenumbers $k_{00,r}^j$ ($+$),
$k_{01,r}^j$ ($\vartriangle$), $k_{02,r}^j$ ($*$) for $L=1m$ and
$b=0.1m$. Figure \ref{deformee_Re} in Appendix C shows the mode
profiles around the two wavenumbers indicated by an arrow.}
\end{figure}

\begin{figure}[!ht]
\begin{center}
\includegraphics[height=.25\textheight,width=.6\linewidth]{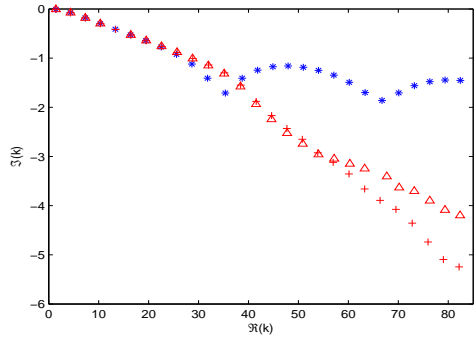}
\end{center}
\caption{\label{zero_0_HM}Resonance wavenumbers $k_{00,r}^j$ of the
first series ($m=0$,$n=0$) with an influence of 0 ($l=0$: $*$), 1
($l=1$: $\vartriangle$) and 2 ($l=2$: $+$) H.M for $L=1m$ and
$b=0.1m$}
\end{figure}

The principal interest of the complex Zorumski's formulation is that
we can observe the influence of higher order duct modes (also
denoted by H.M afterward). In this paper, we taken into account only
the axisymmetric modes ($m=0$) (but the non-axisymmetric are very
easy to calculate with the same method). Figure \ref{zero_3HM} shows
resonance wavenumbers $k_{00,r}^j$, $k_{01,r}^j$, $k_{02,r}^j$. We
observe three series of resonances. Each series starts at the cutoff
frequency of a duct mode. The first one corresponds to a domination
of the planar duct mode, the second one to a domination of duct mode
$01$ and the third one to a domination of duct mode $02$ (see
Appendix B, the Green's function profile is plotted around
$k_{00}^{14}$ and $k_{01}^7$). Fig. \ref{zero_0_HM} shows the
influence of the two first higher order duct modes ($m=0,l=1$ and
$m=0,l=2$) on the resonance wavenumbers of the first series. It is
worth noting that below the first cut off wavenumber
$k_{cut01}=3.83/b$, only one higher order duct mode is sufficient to
accurately describe the resonances (see some values in Table
\ref{table1}); between the first and second cut off wavenumber
$k_{cut02}=7.02/b$, only two higher order duct modes are enough and
similarly to the higher order duct modes: between the $n^{th}$ and
$(n+1)^{th}$ cut off, only $(n+1)$ higher order duct modes are
enough.

\begin{table}[!h]
\begin{center}
\begin{tabular}{cccc}
  $j$ & $k_{00}^j$ & $k_{00,r}^{j}$ with 0 H.M & $k_{00,r}^{j}$ with 1 H.M  \\
  \hline
  0 & 1.5708 & 1.449 - 0.0095i & 1.451 - 0.0096i \\
  1 & 4.712 & 4.369-0.077i & 4.375-0.0776i  \\
  2 & 7.854 & 7.333-0.182i & 7.345-0.183i  \\
  3 & 10.996 & 10.336-0.296i & 10.345-0.299i  \\
  4 & 14.137 & 13.365-0.412i & 13.4-0.415i  \\
  5 & 17.279 & 16.409-0.526i & 16.45-0.53i  \\
  6 & 20.42 & 19.463-0.645i & 19.523-0.644i  \\
  7 & 23.562 & 22.523-0.773i & 22.608-0.76i  \\
  8 & 26.7 & 25.59 -0.924i & 25.707-0.881i \\
  9 & 29.85 & 28.674-1.12i & 28.824-1.01i  \\
  10 & 32.99 & 31.822-1.408i & 31.965-1.152i  \\
  11 & 36.13 & 35.377-1.711i & 35.149-1.317i  \\
  12 & 39.27 & 38.747-1.409i & 38.395-1.579i   \\
\end{tabular}
\end{center}
\caption{\label{table1}Values of the $j$ first resonance wavenumbers
without radiation ($k_{00}^j$) and with radiation ($k_{00,r}^{j}$)
for 0 and 1 H.M, with $b=0.1 m$ and $L=1m$.}
\end{table}

\newpage
\subsection{Evolution of the $j$ first resonance wavenumbers with respect to radiation}
In the present section, we show the evolution of the $j$ first
resonance wavenumbers when only the planar duct mode propagates with
respect to radiation and as shown in the previous section, we take
into account the effect of one higher order duct mode. For this
purpose, we introduce a multiplicative coefficient on the radiation
impedance. Physically, this coefficient $\eta_{\rho}$ can be
regarded as the ratio between the external fluid density
$\rho_{ext}$ and the internal fluid density $\rho_{int}$, such as:
\begin{equation}\label{eta_rho}
\eta_{\rho}=\frac{\rho_{ext}}{\rho_{int}},
\end{equation}
the density of the external fluid $\rho_{ext}$ varying from a vacuum
to water density, the sound celerity, $340 m.s^{-1}$, remaining
constant.

\begin{figure}[!ht]
\begin{center}
\includegraphics[height=.25\textheight,width=.6\linewidth]{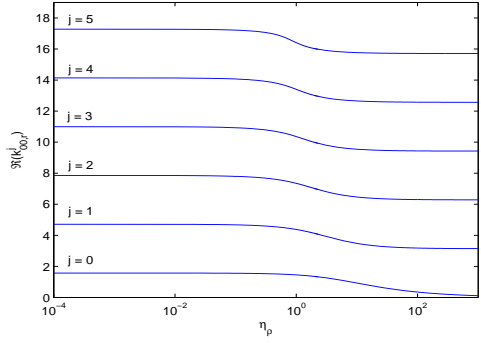}
\end{center}
\caption{\label{j_long_re} Evolution of the real part of the $j$
first longitudinal resonance wavenumbers $k_{00,r}^{j}$ with respect
to $\eta_{\rho}$, for $L=1m$ and $b=0.1m$.}
\end{figure}

\begin{figure}[!ht]
\begin{center}
\includegraphics[height=.25\textheight,width=.6\linewidth]{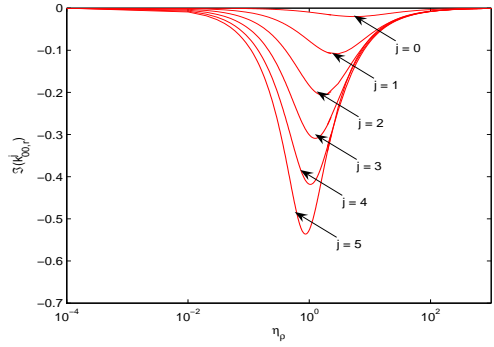}
\end{center}
\caption{\label{j_long_im} Evolution of the imaginary part of the
$j$ first longitudinal resonance wavenumbers $k_{00,r}^{j}$ with
respect to $\eta_{\rho}$, for $L=1m$ and $b=0.1m$.}
\end{figure}

Figures \ref{j_long_re} and \ref{j_long_im} show that when the
parameter $\eta_{\rho}$ increases, a Neumann/Neumann problem is
obtained: the real
 part tends to $j\pi /L$ and the imaginary part tends
  to zero. This behavior corresponds to a system without losses, the external
   fluid becoming a perfectly reflective surface.\\
Fig. \ref{j_long_im} shows that the absolute value of the imaginary
part of the resonance frequency goes through a maximum,
corresponding to the maximum of the radiation. Similarly results
have been observed for non planar modes. So, we can conclude that
energy radiation losses evolves
with the densities of internal and external fluid and goes through a maximum for a specific densities ratio.\\

\section{Conclusion} \label{conclusions}
A development of the Green's function for the Helmholtz equation in
a free space valid for complex frequencies is possible and leads to
a new formula in Zorumski's radiation impedance. The interest has
been shown for an application example, dedicated to the calculations
of the complex resonance frequencies of a radiating flanged
cylindrical duct. It has been shown that length correction
calculated for a semi-infinite duct is a good estimate of a finite
duct radiation when the ratio $L/b \geq 2$ and for frequencies
sufficiently below the first cut off frequency. The study of the
influence of higher order duct modes has shown that below the first
cut off frequency, only one higher order duct mode is needed to
accurately describe the influence of the external fluid on the
resonances and between the $n^{th}$ and $(n+1)^{th}$ cut off, only
$(n+1)$ higher order duct modes are enough. In the last part, it has
been observed a maximum of radiation for a specific densities ratio.
In the future, it will be interesting to use a BEM method to study
more complicated geometries and to observe the resonances with an
experimental method. This work is a first step to study the relation
between radiation and several parameters in order to optimize
geometry for minimizing (e.g. for noise pollution) or maximizing the
sound radiation (e.g. for wind instruments).\\

\section*{Acknowledgements}
The authors wish to thank P. Herzog and F. Silva for their useful
discussion points.

\appendix

\section{\label{Appendix A}Matricial calculation}

The main steps required to obtain the results of section \ref{sec:method of cascade impedances} are presented here (see Ref. \cite{kergo_transmi}).\\
\indent For a cylinder, the general solutions at point $z_1$ can be
described with respect to the values of P and V at point $z_2$ such
as described by relation (\ref{trans}). For all the modes $mn$, the
following matrix problem is obtained:
\begin{equation}\label{Pz0_a}
\mathbf{P}(z_1)=\mathbf{C}\mathbf{P}(z_2)+\mathbf{Z_c}\mathbf{S}\mathbf{V}(z_2),
\end{equation}
\begin{equation}\label{Uz0_a}
\mathbf{V}(z_1)=\mathbf{Z_c^{-1}}\mathbf{S}\mathbf{P}(z_2)+\mathbf{C}\mathbf{V}(z_2),
\end{equation}
$\mathbf{C}$ being a diagonal matrix constituted by the elements
$\cosh(ik_{mn}(z_2-z_1))$ et $\mathbf{S}$ a diagonal matrix
constituted by the elements $\sinh(ik_{mn}(z_2-z_1))$. Eq.
(\ref{Uz0_a}) implies:
\begin{equation}\label{Pz0l_a}
\mathbf{P}(z_2)=\mathbf{Z_cS^{-1}}\mathbf{V}(z_1)-\mathbf{Z_cS^{-1}}\mathbf{C}\mathbf{V}(z_2).
\end{equation}

Introducing Eq. (\ref{Pz0l_a}) in Eq. (\ref{Pz0_a}) and using the
commutativity of the diagonal matrices, we obtain:
$$\mathbf{P}(z_1)=\mathbf{Z_cS^{-1}CV}(z_1)-[\mathbf{Z_cS^{-1}CC}-\mathbf{Z_cS}]\mathbf{V}(z_2),$$
with
\begin{eqnarray}\label{CC_a}
\mathbf{Z_cS^{-1}CC}-\mathbf{Z_cS}
&=&\mathbf{Z_cS^{-1}}(\mathbf{I}+\mathbf{SS})-\mathbf{Z_cS}
\nonumber\\
&=&\mathbf{Z_cS^{-1}}+\mathbf{Z_cS}-\mathbf{Z_cS}
\nonumber\\
&=&\mathbf{Z_cS^{-1}}, \nonumber
\end{eqnarray}

thus:
\begin{equation}\label{Pz0final_a}
\mathbf{P}(z_1)=\mathbf{Z_cS^{-1}CV}(z_1)-\mathbf{Z_cS^{-1}V}(z_2).
\end{equation}

Therefore, with Eqs. (\ref{Pz0l_a}) and (\ref{Pz0final_a}), we
obtain the relation (\ref{multiporte}).

\section{\label{Appendix B}Green's function profile in the duct around resonance frequencies}

\begin{figure}[!h]
\begin{center}
\includegraphics[height=.4\textheight,width=.8\linewidth]{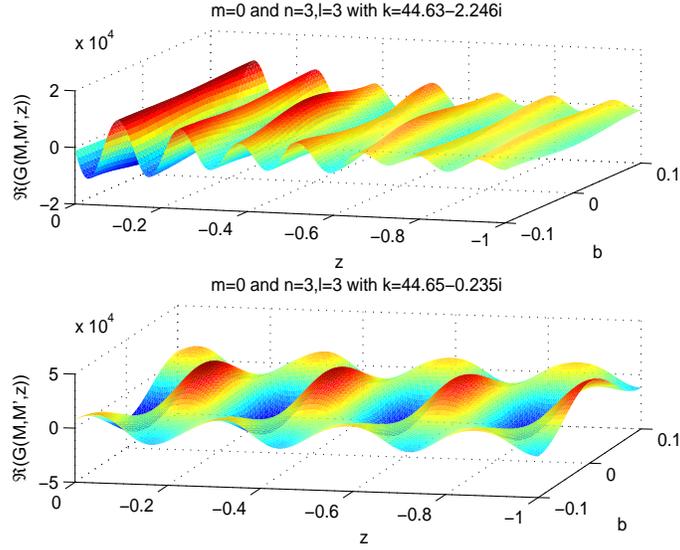}
\end{center}
\caption{\label{deformee_Re}Profile in the duct of the real part of
Green's function around $k_{00}^{14}$ with 3 H.M (upper figure) and
around $k_{01}^{7}$ with 3 H.M (lower figure)}
\end{figure}

Figure \ref{deformee_Re} shows that around the resonance frequency
$k_{00}^{14}$, the profile of Green's function corresponds to the
profile of the planar duct mode even if duct mode $01$ is
propagating and similarly around the resonance frequency
$k_{01}^{7}$, the profile of Green's function corresponds to the
profile of the first non planar duct mode even if planar duct mode
is propagating. The same comportment is observed around other
resonance frequencies. Therefore, it is worth noting that each
series observed in Fig. \ref{zero_3HM} corresponds to a predominant
duct mode even if other duct modes are propagating. Notice that the
evanescent duct modes exist mainly near to the source (at $z=-0.5$):
this clearly appears on the upper figure.

\newpage
\bibliographystyle{elsarticle-num}
\bibliography{article_jasa}
\end{document}